# Fluorescence decay data analysis correcting for detector pulse pile-up at very high count rates

**Matthias Patting, Paja Reisch, Marcus Sackrow, Rhys Dowler, Marcelle Koenig, Michael Wahl**[*]

PicoQuant GmbH, Rudower Chaussee 29, 12489 Berlin, Germany

**Abstract**. Using Time-Correlated Single Photon Counting (TCSPC) for the purpose of fluorescence lifetime measurements is usually limited in speed due to pile-up. With modern instrumentation this limitation can be lifted significantly but some artefacts due to frequent merging of closely spaced detector pulses (detector pulse pile-up) remains an issue to be addressed. We propose here a data analysis method correcting for this type of artefact and the resulting systematic errors. It physically models the photon losses due to detector pulse pile-up and incorporates the loss in the decay fit model employed to obtain fluorescence lifetimes and relative amplitudes of the decay components. Comparison of results with and without this correction show a significant reduction of systematic errors at count rates approaching the excitation rate. This allows quantitatively accurate fluorescense lifetime imaging (FLIM) at very high frame rates.

**Keywords**: fluorescence lifetime, photon counting, TCSPC, data analysis, fitting, FLIM, FRET.

**\*** E-mail: wahl@picoquant.com

## 1 Introduction

Time-Correlated Single Photon Counting (TCSPC) is one of the most important methods to determine fluorescence lifetimes on the nanosecond scale [O'Con 84]. It allows higher sensitivity and better time resolution than most analog detection methods and is well established also for fluorescence lifetime imaging (FLIM) in confocal microscopy for many important life science applications [Lako 06].

In TCSPC one repeatedly measures and histograms the elapsed time between sample excitation by a pulsed light source and the arrival of emitted fluorescence photons at the detector, typically with picosecond accuracy. However, classic TCSPC electronics as well as most single photon counting detectors have dead times on the order of several tens of nanoseconds. Within this time the system is busy with data processing and cannot detect any other photon. The dead time is typically longer than the fluorescence decay processes of interest. Consequently, if more than one photon is emitted per excitation cycle it cannot be detected. This causes a statistical over-representation of early photons and a corresponding distortion of the collected decay shape, an effect called pile-up [O'Con 84, Kapu 15]. In order to avoid this situation it is necessary to work with sufficiently low light intensities. Due to the statistical nature of photon flux this results in the unfortunate situation that most of the excitation pulses lead to no emission at all and the speed of photon collection is reduced to typically one out of every 50-100 laser pulses [O'Con 84, Kapu 15].

In order to acquire a FLIM image with a confocal microscope, one typically scans the excitation and detection foci across the sample [Kobe 03, Ortm 04]. In case of a fast scanning system, this very often leads to only a few detected photons per pixel, which will not allow accurate lifetime analysis. It is therefore necessary to increase the number of detected photons per pixel either by reducing the scan speed or by accumulating several frames to sum up the photons. This is why it typically takes several seconds up to half a minute to obtain a good FLIM image. A possible path to speedier FLIM is to perform TCSPC with the highest possible photon rate.

Even though it is feasible to correct for dead-time [Patt 07, Isba 17] the ultimate key to high throughput TCSPC is eliminating dead-time. However, the highest time resolution and lowest differential nonlinearity (DNL) in timing can typically only be obtained with dedicated time-to-digital converters that incur a dead-time on the order of some tens of nanoseconds after each photon detection. It is possible to build faster timing circuits that allow effectively zero dead-time but compromise on timing resolution and DNL. We have recently developed a TCSPC board that achieves 40 Mcps throughput, a dead-time < 1 ns, and a resolution of 250 ps, without compromising on DNL [Pico 17]. Using such a TCSPC unit with negligible dead-time it is possible to detect several photons within one excitation cycle. Although the temporal resolution of the current hardware design is only 250 ps, it is still sufficient for most FLIM applications in the life sciences. In addition to the timing electronics, the detector must also meet the requirement of short dead time. Currently the most suitable detectors in this regard are so called hybrid photo detectors (HPD) based on a combination of two gain stages, a first one similar to that of a photomultiplier tube and a second one in the form of an avalanche photo diode [Hama 07]. With this combination of TCSPC unit and detector we have been able to demonstrate TCSPC data collection speeds of up to 40 Mcps at excitation rates of 40 MHz, which is about 100 times faster than with conventional TCSPC and conservative pile-up constraints. Even when compared with relaxed pile-up constraints (allowing more error, which is sometimes considered tolerable in FLIM) a speedup by a factor of 10 is obtained. This speeds up FLIM acquisition accordingly, provided that the sample emits enough light.

One issue that remains with the approach sketched so far is that of detector pulse pile-up. This effect is caused by the fact that individual detector pulses have a certain width that cannot be made infinitely small. With existing hardware as introduced above this is about 500 ps. At the desired high count rates the actual photon statistics imply that there will quite frequently be photon emissions closer than this, which means that successive detector pulses may overlap and merge into one. Since the TCSPC electronics cannot distinguish the original pulses they will be counted as only one. This causes another form of histogram distortion where early photons are under-represented. In the following we introduce a correction scheme for the systematic errors resulting from this type of distortion. More precisely, we show how the underlying physical mechanism of photon losses due to pulse pile-up can be incorporated in the fit model so that the fluorescence lifetime estimate is made robust against this type of histogram artefact, even at very high count rates.

## 2 Proposed Correction Metod

A key parameter for the photon loss model is the closest pulse spacing in time that the particular combination of TCSPC electronics and detector can still resolve. Let this pulse pair resolution be called $\Delta t$ here. Assuming Poisson statistics and a count rate n, the time intervals between successive photons will be exponentially distributed and the fraction of intervals shorter than $\Delta t$ will be $1 - e^{-n\Delta t}$. A photon falling into $\Delta t$ after the detection of a previous photon will be lost. The fraction of remaining photons will therefore be $e^{-n\Delta t}$ [Reed 72].

TCSPC with pulsed excitation samples the photon distribution as a function of elapsed time t after excitation. Let Dec(t) be the decay curve. With known excitation rate $f_{exc}$ and known pulse count N each discrete time $t_i$ can be associated with a „differential" or instantaneous count rate

$$dn = f_{exc} \, Dec(t_i) / N . \qquad (1)$$

Applying Reed et al [Reed 72] for each time $t_i$, one observes a lossy decay curve

$$\mathrm{Dec_{Exp}}(t_i) = \mathrm{Dec}(t_i)\, e^{-d_n\, \Delta t} = \mathrm{Dec}(t_i)\, e^{-f_{exc}\, \mathrm{Dec}(t_i)/N\, \Delta t}. \qquad (2)$$

The losses scale exponentially and distort the shape of the curve accordingly. Attempting to fit this observed decay curve with a model function $\mathrm{Dec_{Mod}}$ via nonlinear least squares methods (or maximum likelihood estimation) in order to obtain physical parameters such as fluorescence lifetimes will result in systematic errors of the parameters.

It is possible to correct the observed decay curve iteratively and thereby reconstruct the true decay curve $\mathrm{Dec}(t)$, as shown previously for classic pile-up [Isba 16]. Unfortunately the procedure is time consuming and would distort the experimental noise of the counting statistics. This is undesirable since the well defined Poisson statistics of the experimental error are of critical importance in the process of maximum likelihood estimation when fitting a model to the observed data.

The idea of the correction method proposed here is not to correct the observed decay curve but to incorporate the losses according to Eq. 2 into the fitted model function. This results in a modified model curve

$$\mathrm{Dec_{Corr}}(t) = \mathrm{Dec_{Mod}}(t)\, e^{-f_{exc}\, \mathrm{Dec_{Mod}}(t)/N\, \Delta t}. \qquad (3)$$

This model curve is fitted to the data as usual by means of a nonlinear least squares algorithm or maximum likelihood estimation. The pulse pair resolution $\Delta t$ can either be determined in advance as an instrument specific constant or alternatively be obtained as an additional fit parameter. The latter allows a calibration measurement of Δt to be performed in reverse without additional expenditure of apparatus or method. Nevertheless it must be noted that despite the correction, the loss of photons due to detector pulse pile-up will cause some loss in accuracy of the curve fitting and parameter estimation. These losses in accuracy, however, emerge only as statistical errors, which is a substantial improvement over the uncorrected situation, where systematic errors dominate over statistical errors.

A visual explanation of the proposed correction method is shown schematically in Fig. 1. For simplicity it uses a single-exponential decay and for clarity the photon losses due to detector pulse pile-up are exaggerated.

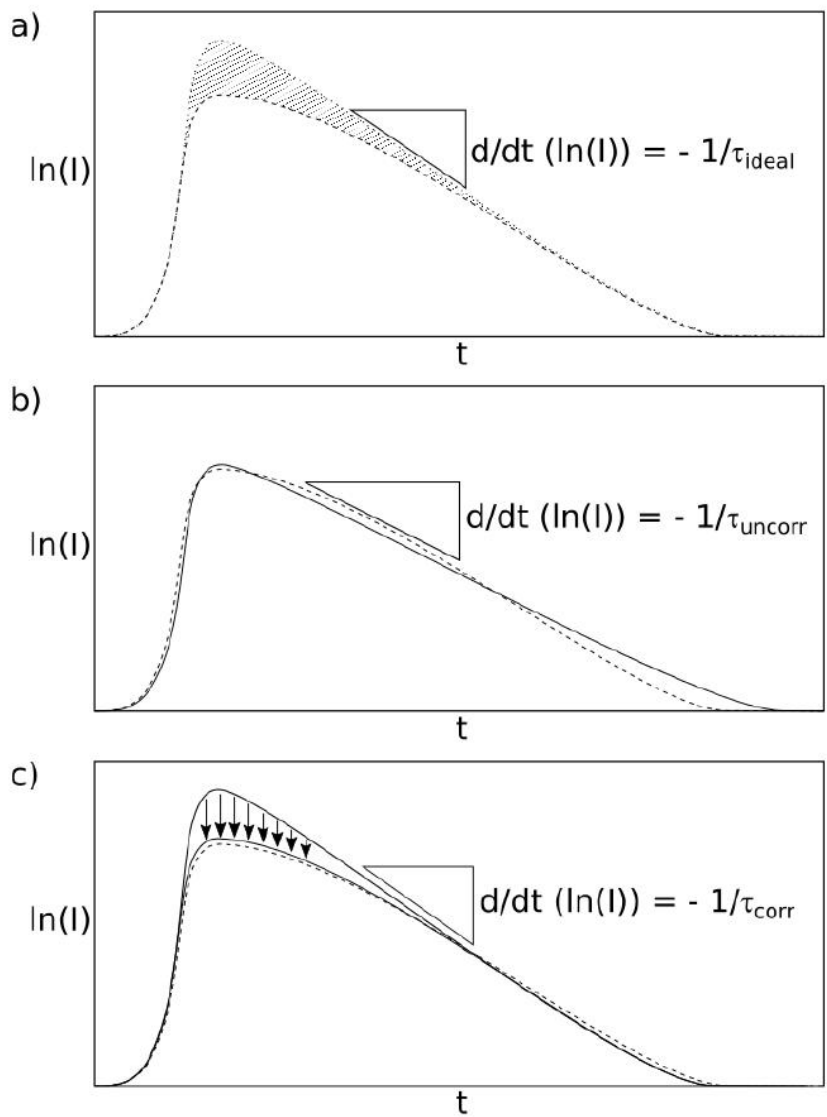

**Fig. 1** Schematic illustration of the correction scheme by example of a single-exponential decay

a) The original undistorted decay has a lifetime $\tau_{ideal}$, symbolized by the slope triangle d/dt (ln(I)). Photon losses due to detector pulse pile-up (hashed area) lead to a compressed peak region in the observed decay (dashed line). Only the latter is accessible by measurement.

b) a nonlinear fit of the observed decay (dashed line) with an uncorrected model function results in lifetime $\tau_{uncorr}$. The deformation results in a longer lifetime than expected.

c) Adjustment of the model function (symbolized by arrows) ensures that the model curve (solid line) is in agreement with the observed decay curve (dashed line). The resulting lifetime $\tau_{corr}$ is now in agreement with the expected value $\tau_{ideal}$.

## 3   Experimental Results

In order to assess the effectiveness of the proposed correction scheme we performed FLIM measurements on a mixture of two types of polymer microspheres stained with dyes of different lifetimes. The sample was prepared by mixing 20 µl Dragon Green beads (Bangs Laboratories Inc., Fishers, IN, USA), 20 µl Nile Red beads (Spherotech Inc., Lake Forest, IL, USA) and 0.5 ml of purified water. A 2 µl droplet of the mixture was then dried on a glass cover slip. Imaging was performed on a MicroTime 200 confocal microscope (PicoQuant, Berlin, Germany). The sample was excited by a pulsed diode laser at 485 nm and 20 MHz pulse rate (PicoQuant GmbH, Berlin, Germany). Emission was separated out via a filter LP488 (Semrock Inc., Rochester, NY, USA) and dichroic zt488/640rpc (Chroma Technology Corp, Bellows Falls, VT, USA). The photon detector was a PMA-Hybrid 40 and the TCSPC electronics were a TimeHarp 260 N (both PicoQuant GmbH, Berlin, Germany). Figure 2 shows the uncorrected and corrected fitting results for the two lifetimes in the brightest pixel of the image (80x80 µm, 256x256 pixels). The calculated lifetimes of the two bead types are plotted against the count rate.

As the beads were spatially well separated (meaning that all pixels exhibited a mono-exponential behavior), a mono-exponential reconvolution model was applied. The lifetimes were taken from their corresponding peak positions in the lifetime histogram of the image, whereas the error bars indicate the FWHM of the lifetime peak. The uncorrected fits (solid square) show a strong dependency on the count rate, whereas the corrected fits (open triangle) stay constant within the tolerance level, as indicated by the dotted line.

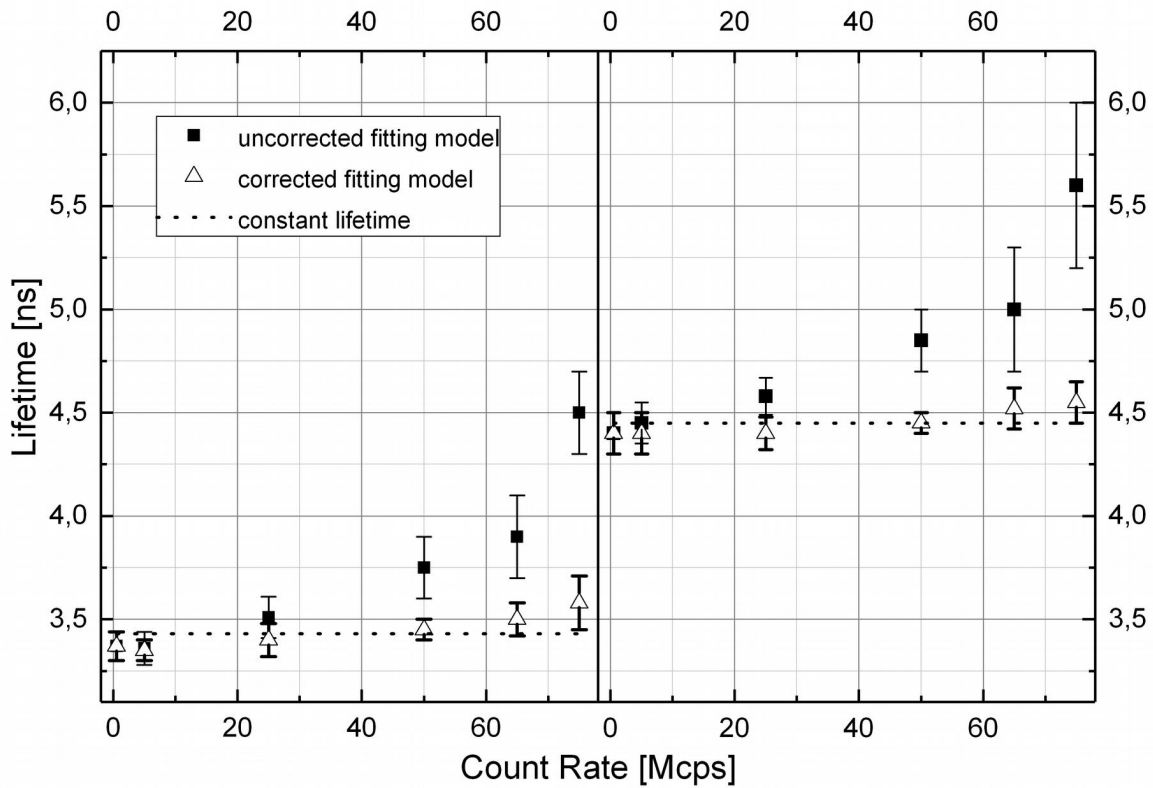

**Fig. 2** Fitting results for the brightest pixel of the image (80x80 µm, 256x256 px) of a bead sample (PMA-Hybrid 40, TimeHarp 260 N, 20 µl Dragon Green beads + 20 µl Nile Red beads + 0,5 ml water, 2 µl droplet dried on glass cover slip, filter: LP488, dichroic: zt488/640rpc, excitation at 485nm and 20 MHz rate). The calculated lifetimes of the two bead types (left and right panel, respectively) are plotted against the count rate.

Note that the width of the uncorrected lifetime peaks shows a pronounced increase with increasing count rate. The reason lies in the intensity gradient along each individual bead (bright center, dark rim). In the center the lifetime distortions by the detector pulse pile-up are strongest, whereas at the rim the effects are less pronounced. For the corrected fits the width of the lifetime peaks stays small and constant, a further indication that the correction helps to reduce the detector pulse pile-up effects.

Above 60 Mcps there appears to be a slight trend towards systematically increased lifetimes even for the corrected lifetimes, indicating that the correction starts to become less precise. The reason for this lies in count rate dependent systematic errors in the decay curves that probably originate from the front end analog electronics or the detector and cannot be described in terms of detector pulse pile-up. Nevertheless, the relative error associated with the trend is still smaller than 5%.

In order to demonstrate the effectiveness of the correction also in Förster Resonant Energy Transfer (FRET) imaging with typically more complex decay shapes we have analyzed three data sets from chinese hamster ovary (CHO) cells with EGFP-N-WASP and mRFP-Toca-1 (courtesy of S. Ahmed and T. Sudhaharan, Institute of Medical Biology, Singapore) with and without correction. Imaging was performed at average count rates of 5, 20, and 30 Mcps on a MicroTime 200 confocal microscope (PicoQuant, Berlin, Germany). The excitation wavelength was 485 nm at 40 MHz repetition rate and emitted photons were collected between 500 - 540 nm using a PMA Hybrid 40 detector module and a TimeHarp 260 N TCSPC unit (both PicoQuant, Berlin, Germany). TCSPC histograms of each pixel were fitted with a double exponential model. Average lifetime results mapped to color are shown in Figure 3. The corrected version is free of visible artefacts up to 20 MHz count rate. At 30 MHz a trend towards falsely long lifetimes is visible but lifetime contrast is still substantially better than uncorrected.

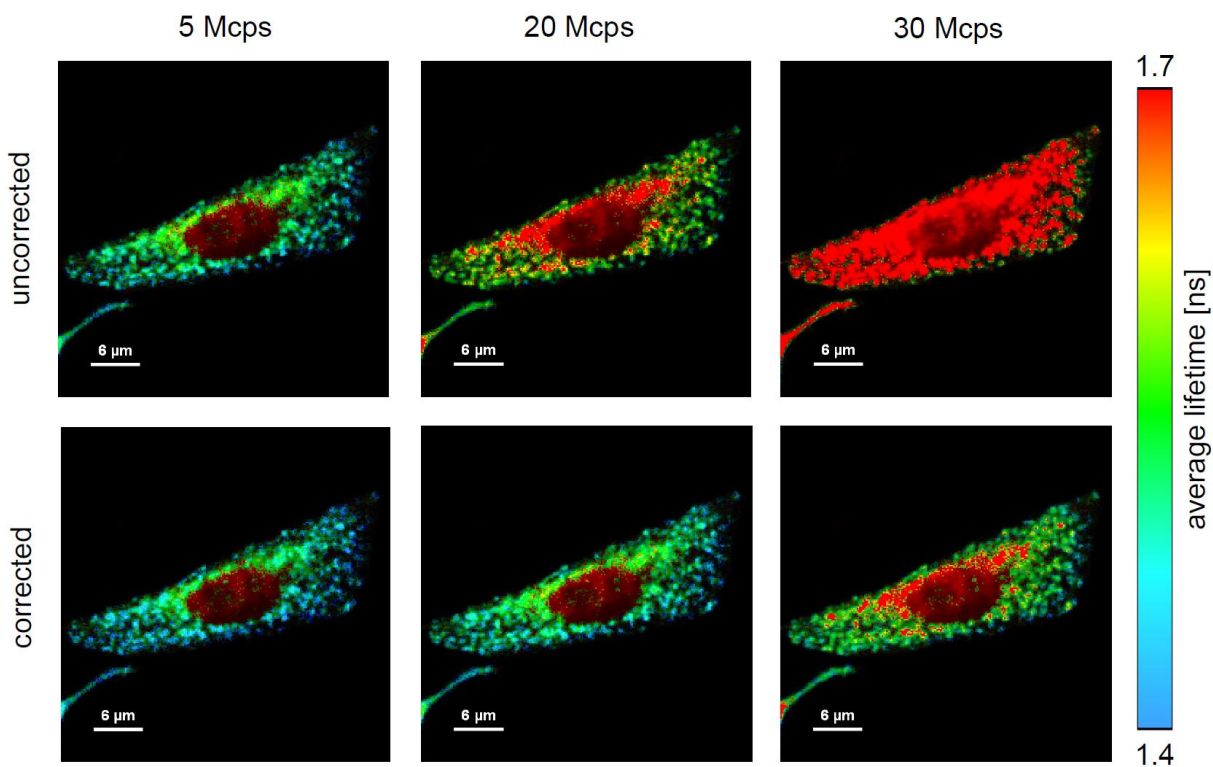

**Fig. 3** FRET imaging results with and without correction at different count rates. Sample: CHO cells with EGFP-N-WASP and mRFP-Toca-1, courtesy of S. Ahmed and T. Sudhaharan, Institute of Medical Biology, Singapore.

## 4   Conclusion

We have shown a correction method for detector pulse pile-up in fluorescence lifetime measurements with TCSPC that allows for data collection at very high count rates. Experimental results indicate that this allows measuring at count rates approaching the excitation rate with lifetime errors below the 5% level. This facilitates quantitatively accurate confocal FLIM measurements at very high frame rates.

## 5 Acknowledgements

The authors thank Sandra Orthaus-Müller (formerly PicoQuant GmbH) for pushing and inspiring the project. This work was partially funded by BMBF grant 13N12672 "tCAm4Life".